 \documentclass[journal]{IEEEtran}
\usepackage[final]{graphicx}
\usepackage{amssymb,amsmath,amsthm}
\usepackage{chngcntr}

\theoremstyle{definition}

\usepackage{algorithm, algorithmic}
\usepackage{booktabs}
\usepackage{multirow}
\usepackage{bigstrut}
\usepackage{xcolor}
\usepackage{setspace}
\usepackage{cite}

\title{A rolling-horizon dynamic programming approach for collaborative caching}

\author{Xinan Yang and Nikolaos Thomos~\IEEEmembership{Senior~Member,~IEEE}
\thanks{X. Yang is with the department of Mathematical Sciences, University of Essex, Colchester, United Kingdom (e-mail: xyangk@essex.ac.uk). N. Thomos is with the School of Computer Science and Electronic Engineering, University of Essex, Colchester, United Kingdom (e-mail: nthomos@essex.ac.uk).} 
}

\begin{document}

\maketitle

\begin{abstract}
In this paper, we study the online collaborative content caching problem from network economics point of view. The network consists of small cell base stations (SCBSs) with limited cache capacity and a macrocell base station (MCBS). SCBSs are connected with their neighboring SCBSs through high-speed links and collaboratively decide what data to cache.
Contents are placed at the SCBSs ``free of charge'' at off-peak hours and updated during the day according to the content demands by considering the network usage cost. We first model the caching optimization as a finite horizon Markov Decision Process that incorporates an auto-regressive model to forecast the evolution of the content demands. The problem is NP-hard and the optimal solution can be found only for a small number of base stations and contents. To allow derivation of close to optimal solutions for larger networks, we propose the \textit{rolling horizon} method, which approximates future network usage cost by considering a small decision horizon. The results show that the \textit{rolling horizon} approach outperforms comparison schemes significantly. Finally, we examine two simplifications of the problem to accelerate the speed of the solution: (a) we restrict the number of content replicas in the network and (b) we limit the allowed content replacements. The results show that the \textit{rolling horizon} scheme can reduce the communication cost by over 84\% compared to that of running Least Recently Used (LRU) updates on offline schemes. The results also shed light on the tradeoff between the efficiency of the caching policy and the time needed to run the online algorithm.
\end{abstract}


\section{Introduction}
During the last few years, we have witnessed an explosion of the data traffic in cellular networks. Due to the increase in the number of wireless devices with network access, the mobile video traffic will reach 82\% of the overall Internet traffic by 2021 \cite{Cisco}. This increase puts pressure on network operators infrastructure and renders inefficient the current model according to which the base stations (BS) receive the requested content through the core network using expensive bandwidth-limited backhaul links. The backhaul links may become congested because of the increased data traffic, and this may result in users experiencing excess delays and low Quality of Experience (QoE). High QoE can be preserved by densifying the network of BSs, \textit{i.e.}, installing a larger number of BS; however, this solution does not scale well with the number of wireless devices. A more efficient solution would be to exploit the spatial diversity of the data requests. This can be done by caching popular content at BSs' caches so that it is closer to the end-users. The exploitation of BSs' caches helps to decrease the load of the backhaul links and combat download delays.

Edge caching resembles web caching mechanisms employed in Content Delivery Networks (CDN), where content providers replicate popular content in dedicated servers (mirror servers), which accommodate users' requests. As the content is located near the end-users, the communication with the origin servers is limited. Thus, caching helps minimize the data traffic in the core network. The CDN approach of having an increased number of servers to accommodate the demands of the users does not scale well with the number of users and leads to increased delays and low QoE \cite{VakaliIntComp2003}. Efficient CDN architectures \cite{Akamai} optimize the cached content in the mirror servers so that the load of the core network is reduced. In the wireless caching systems, the cache optimization problem is harder than that in CDNs due to the plethora of the requested contents and the dynamic nature of the requests. This is fueled by the change of the content production and consumption model which has transformed the end-users to content producers. This problem becomes even more challenging as the vast majority of the generated data is video, which has considerable size and is characterized by strict delivery deadlines. Hence, each base station can cache only a part of the entire content catalogue. The cache optimization problem can be mapped to a knapsack problem which is known to be NP-hard \cite{KnapsackProbBook}. In practice, network operators adopt simple cache update policies such as the Least Recently Used (LRU) or the Least Frequently Used (LFU).

Collaborative wireless caching is proposed in \cite{GolrezaeiCommMag2013,GolrezaeiTITApr2014} in order to take advantage of the spatial correlation of the requests. Specifically, the BSs collaborate with wireless devices which offer their caching space (helper nodes) to store popular content. 
The advantages of coded caching \cite{Shokrollahi06,AhlswedeTIT00} are explored in \cite{GolrezaeiCommMag2013,MaddahAliTIT2014} where the benefits of content reuse in increasing the cache hit ratio and decreasing the content delivery delay are shown. More recently, the tradeoff between global caching gains and pre-downloading gains is investigated in \cite{GregoriJSAC2016} for the aforementioned network setting. In \cite{PoularakisTCOM2014,PoularakisInfocom16,KhreishahJSAC2017}, a large number of cache-enabled small-cell base stations (SCBSs) are placed to lower the data retrieval delays in mobile networks. The cached content in the SCBSs is decided centrally by the mobile network operator (MNO). The SCBSs work in concert with a macro-cell base station (MCBS) to deliver content that is not available to the SCBSs, which is requested from MCBS. In \cite{PoularakisTCOM2014}, the problem is first cast as an unsplittable hard capacitated metric facility location problem and then is solved efficiently by an approximation algorithm which performs close to the optimal solution. Cache assisted delivery of scalable video data \cite{ThomosTMM2015} is studied in \cite{PoularakisInfocom16} where multiple MNOs collaborate locally by offering part of their cache space to SCBSs of other MNOs. It is shown that the cache optimization is related to the multiple-choice knapsack problem and that a pseudo-polynomial time-optimal solution exists. The relation between partial caching and users retention rate for video data is investigated in \cite{MaggiCacheComCom2018}. It is suggested that only the initial part of the video files should be cached in the BSs and that the size of the cached part depends on the video popularity. 
In \cite{KhreishahJSAC2017}, a joint caching, routing, and channel assignment scheme is presented for optimizing video delivery over coordinated small-cell cellular systems. The column generation method is used to reduce the computational complexity and to find a well-performing solution. 

The schemes mentioned above assume that the content popularity profile is known and can be modeled as Zipf, Zipf-Mandelbrot, Shot Noise Model, or are a combination of several distributions \cite{FamaeyJNCA2013}. In these schemes, the content replacement decisions are made assuming that the future content requests will be in accordance with the considered model. Hence, offline schemes depend heavily on the sufficiency of content request statistics that are used for parameter fitting and are sensitive to sudden changes in popularity. When content popularity is time-varying, offline caching schemes perform sub-optimally. To address the inefficiency of offline schemes for non-stationary content demands online algorithms have been proposed \cite{LiTMM17,ZhangICCW18, YangArxiv18, AbadArxiv18, SadeghiJTSP18, NegliaToN18, BharathTCOM18, ChattopadhyayTWC18,GharaibehInfocom16,MullerTWC2017,SaltarinNetworking2018}. There exist both non-collaborative \cite{ZhangICCW18, YangArxiv18, AbadArxiv18, SadeghiJTSP18, NegliaToN18, BharathTCOM18,MullerTWC2017,LiTMM17} and collaborative caching schemes \cite{ChattopadhyayTWC18,GharaibehInfocom16,MullerTWC2017,SaltarinNetworking2018}. Base stations in non-collaborative caching decide independently from each other the optimal caching policy. To determine the optimal online cache policy reinforcement learning algorithms have been suggested \cite{ZhangICCW18, AbadArxiv18, SadeghiJTSP18,MullerTWC2017}. When the freshness of the data is important, caching algorithms should consider the age of information \cite{ZhangICCW18,AbadArxiv18}. More efficient caching policies can be found by taking into account both the global and local content popularity \cite{SadeghiJTSP18}. In \cite{SadeghiJTSP18}, linear function approximation is used in an attempt to lower the computational cost, which is inspired by the additive form of the overall cost. This allows to solve larger instances of the problem, e.g., more files. Context-aware caching is proposed in \cite{MullerTWC2017} for proactively deciding the cache allocation strategy. The content popularity is learned using a contextual multi-armed bandit algorithm which has guaranteed convergence, however, this algorithm still requires a significant number of iterations to learn data popularity. A linear regression model is used to estimate the content popularities in \cite{YangArxiv18}. The problem is first formulated as a time-averaged hit rate maximization problem and then reformulated as time-averaged regret minimization. Two algorithms based on simulated annealing are introduced in \cite{NegliaToN18} to deal with non-stationary content requests. In \cite{BharathTCOM18}, caches are updated when the offloading cost, i.e., the cost introduced when files are delivered through the backhaul, exceeds a threshold. An online popularity-aware caching scheme that refreshes the cached content according to the revealed content requests is presented in \cite{LiTMM17}. This scheme considers a single cache network, and thus the algorithm is not appropriate for collaborative caches case. 
Exploiting collaboration opportunities between BSs can reduce offloading cost \cite{ChattopadhyayTWC18,SaltarinNetworking2018} and/or reduce the network use \cite{GharaibehInfocom16,GharaibehTPDS2016}. Collaboration opportunities may emerge due to the fact the coverage areas of base stations overlap as in \cite{ChattopadhyayTWC18}, where a Gibbsian based sampling method is used to determine the optimal caching strategy. This enables sequential cache updates and policies are updated only when a content is delivered to a user through the backhaul link. 
Collaboration between caches can also be achieved by means of a content centric networks (CCN). 
Caching in CCN from an economic point of view is explored in \cite{GharaibehInfocom16,GharaibehTPDS2016}. Incentives are given to Internet Service Providers (ISPs) to cache data for other ISPs. The proposed algorithm is of low complexity, however, due to the special characteristics of content demand in the CCN framework, it cannot be trivially generalized to solve wireless cache optimization problems. Network coded cached in CCN is studied in \cite{SaltarinNetworking2018} where more popular contents are cached to the edge nodes, from where the end-users can acquire them. 

In this work, we study the wireless edge cache optimization problem from a single MNO point of view. We assume SCBSs with limited storage space that collaborate with each other to accommodate the content requests. We assume that SCBSs are connected with neighboring SCBSs through the core network via high-speed links. When a user issues a request for a content, if it is not cached in the SCBSs where the user lies, the request is forwarded to its neighboring SCBSs. If one of these SCBSs has the requested content, it transmits the content to the SCBS where the content was requested, otherwise the request is forwarded further. We examine the problem from a network economics point of view, and hence we consider the cost of updating the content (including the communication cost) to decide whether the content should be retrieved from the MCBS or the SCBSs. The above content request model resembles the way information is requested in information-centric networks and peer-to-peer networks, however our solution is generic and applicable to any network following such a content request model. More details regarding the communication model are provided in Section \ref{sec:net_setting}. The considered scenario has similarities with the one examined in \cite{KhreishahInfocomWK2015,GharaibehTMC2016} where centralized offline caching solutions are presented. Different from \cite{KhreishahInfocomWK2015,GharaibehTMC2016}, we propose an online caching solution which belongs to the family of approximate dynamic programming methods called \textit{rolling horizon}. The \textit{rolling horizon} method approximates future network usage cost and enables the determination of close-to-optimal caching policies in large networks. In our caching network, cache content is placed free-of-charge at the SCBSs in off-peak hours and updated during the day according to the encountered content popularity dynamics considering the cost of update. The changes in the content popularity are captured by an auto-regressive model. We compare the performance of the proposed scheme against several schemes with and without prediction. The results show that the proposed \textit{rolling horizon} approach outperforms all other schemes in all cases and that it needs only the consideration of a small horizon. Further, from the results, it is clear that the policy with the lower network usage cost does not coincide with one achieving the maximum cache hit ratio. As the computational complexity of the problem can be still high, we study two simplifications of the problem: (a) we consider that the content can be cached in at most one SCBS and (b) we limit the number of allowed content replacements in each SCBS. The results show that besides these simplifications large gains over the comparison schemes are noticed and the performance is comparable to a theoretical lower bound achieved when we can accurately predict the demand evolution in the future and update the cached contents without any cost. The contributions of our work are summarized as follows: 
\begin{itemize}

\item
we propose a dynamic programming structure with finite horizon to solve the online cache optimization problem;

\item 
we cope with the dimensionality by proposing two simplification problems to the problem, i.e., we restrict to one the number of replicas in the network and we limit the number of content replacements;

\item
we compare the proposed scheme with several schemes to show the efficiency of the proposed online caching algorithm.

\end{itemize}

The rest of the paper is organized as follows. In Section \ref{sec:net_setting} we discuss the considered scenario. Then we present the online cache optimization model in Section \ref{sec:onlinemodel}. Next, in Section \ref{sec:rollinghorizon}, we propose the \textit{rolling horizon} algorithm and some simplifications of it, which permit to reduce the computational complexity and allow us to use this algorithm in larger problems. The performance of proposed schemes is compared against several schemes in Section \ref{sec:results}. Finally, conclusions are drawn in Section \ref{sec:conclusions}.

\section{System model}
\label{sec:net_setting}

\begin{figure}[t]
\centering
\includegraphics[width=0.45\textwidth]{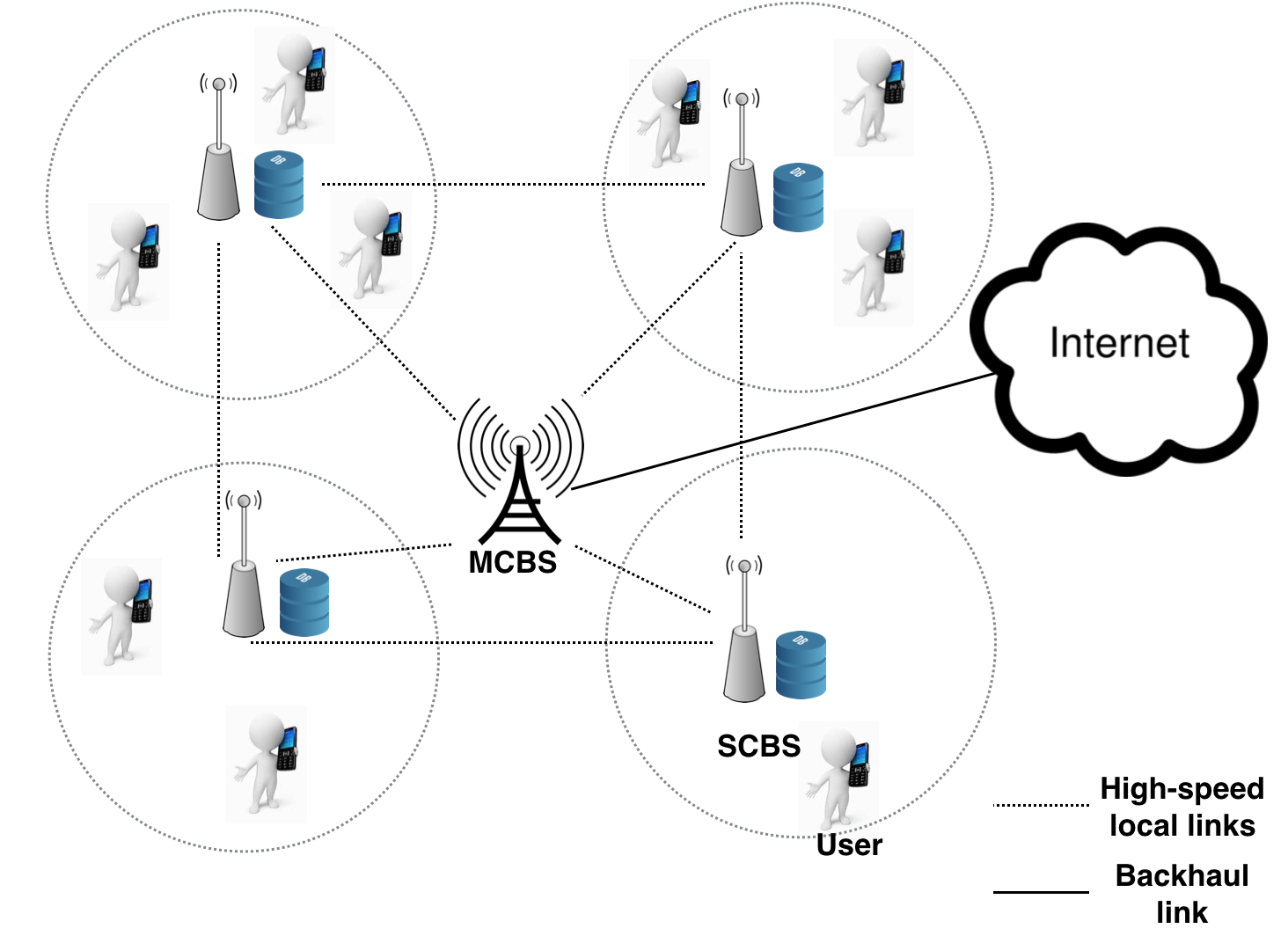}
\caption{Considered network setting.}
\vspace{-0.5 cm}
\label{fig:netset}
\end{figure}

In this paper, we consider the network setting depicted in Fig. \ref{fig:netset} that represents an MNO handled network comprising an MCBS that communicates through the backhaul link with a set $\mathcal{M}=\{1,2, \ldots, M\}$ of SCBSs. Let $\mathcal{M}^{'} = \{ \mathcal{M} \cup 0\}$ be the augment set that includes the MCBS and the SCBSs, where the index $0$ stands for the MCBS. In addition, each SCBS is connected with its neighboring SCBSs through the core network via high-speed links\footnote{The high-speed links can be backhaul links connecting the edge nodes of the core network, i.e., the SCBSs.}. We assume that there is a set $\mathcal{U}=\{1,2, \ldots, U\}$ of users who request to receive files contained in the content catalogue $\mathcal{N}=\{1,2,\ldots,N\}$. Each file $n \in \mathcal{N}$ has size $v_n$ bytes and is associated with a parameter $\lambda_n^t$ that represents the number of requests for the $n$th file in time period $t$. The files are unsplittable, and hence they should be fully retrieved only from one base station (either MCBS or SCBS), however a file $n \in \mathcal{N}$ can be stored in multiple SCBSs. The SCBSs have limited storage capacity, and thus they can cache only a part of the content catalogue, while the MCBS can store the entire content catalogue. The storage capacity of the $m$th SCBS is denoted as $b_m$.

When a user $u \in \mathcal{U}$ issues a request for a content, e.g., $n \in \mathcal{N}$, it first directs this request to the closest SCBS, e.g., $m$th SCBS. In this paper, we assume that contents requests follow the independent reference model (IRM). 
If the content is cached in this SCBS, the delivery of the content incurs no extra cost. 
However, in case the content is not found at the $m$th SCBS, the content request is forwarded to all its neighboring SCBSs. If these do not possess the requested file, they forward it further to their neighbors and so on. When the content is located in the cache of the SCBS $m^{'} \in \mathcal{M}\backslash m$ for example, we assume that the delivery cost is equal to $c_{m^{'}}^u$; this accounts for the communication cost, which includes the cost of using the network infrastructure. In this paper, without loss of generality we assume that the cost depends on the distance of the SCBS $m$, in which the client $u \in \mathcal{U}$ requested the content from, to the resource SCBS (or MCBS), \textit{i.e.}, the $m^{'} \in \mathcal{M}^{'}\backslash m$ where the requested content was located. If none of the SCBSs has the demanded content, this can be acquired from the MCBS through the backhaul link with a cost $c_{0}^u$, where $c_{0}^u > c_{m^{'}}^u$.

\section{Online Dynamic Programming model}
\label{sec:onlinemodel}
In this section, we formulate the studied collaborative online cache optimization problem as a finite-horizon Markov Decision Process (MDP) to reflect the dynamics of the system. We assume that cache decisions are made online and every $T$ decision stages the cached content can be updated free-of-charge, i.e., the cost for updating the cached content at the SCBSs is zero. For example, if we consider $T=24$ and cache updates every hour, the content will be updated free-of-charge at mid-night, i.e., every $24$ hours.

We define the states of the finite-horizon MDP as $S_t = (\vec{x}_1^t, \vec{x}_2^t, ..., \vec{x}_M^t)$, where $\vec{x}_m^t = (x_{1m}^t, x_{2m}^t, ..., x_{Nm}^t)^T$ is a $N$ dimensional $0-1$ vector that indicates whether a content is cached in a SCBS. We have 
\begin{equation}
\begin{split}
x_{nm}^t = \left\{ \begin{array}{ll} 
1, & \mbox{if content $n$ is cached in SCBS $m$ in stage $t$} \\
0, & \mbox{otherwise} \end{array}\right. \\ 
n \in \mathcal{N}, m \in \mathcal{M}. \nonumber
\end{split}
\end{equation}
Recall that $M$ and $N$ represent the number of SCBSs and the number of contents, respectively.

The actions of the finite-horizon MDP are composed of two sub-actions corresponding to the addition and/or the eviction of a content to/from the cache. The sub-action of adding a content to the cache is defined as $\vec{a}_t = (\vec{a}_1^t, \vec{a}_2^t, ..., \vec{a}_M^t)$, where $\vec{a}_m^t = (a_{1m}^t, a_{2m}^t, ..., a_{Nm}^t)^T$ is a $N$ dimensional vector with 
\begin{equation*}
\begin{split}
a_{nm}^t = \left\{ \begin{array}{ll} 
1, & \mbox{ if content $n$ is added in SCBS $m$ in stage $t$} \\
0, & \mbox{ otherwise} \end{array}\right.. \\ 
\end{split}
\end{equation*}
Similarly the sub-action of evicting a content from the cache is defined as $\vec{d}_t = (\vec{d}_1^t, \vec{d}_2^t, ..., \vec{d}_M^t)$, with $\vec{d}_j^t = (d_{1m}^t, d_{2m}^t, ..., d_{Nm}^t)^T$ being a $N$ dimensional vector and 
\begin{equation*}
\begin{split}
d_{nm}^t = \left\{ \begin{array}{ll} 
1, & \mbox{if content $n$ is evicted from SCBS $m$ in stage $t$}\\
0, & \mbox{otherwise} \end{array}\right. \\ 
\end{split}
\end{equation*}
Therefore, the state $S_{t+1}$ and the $x_{nm}^{t+1}$ evolve with the time as follows
$$S_{t} = S_{t-1} + \vec{a}_t - \vec{d}_t.$$
$$x_{nm}^{t} = x_{nm}^{t-1} + a_{nm}^t - d_{nm}^t, n \in \mathcal{N}, m \in \mathcal{M}.$$
For each state $S_t$, the feasible action set for stage $t$, i.e. the set of eligible actions, is given by
$$ \begin{array}{ll} \chi_t = \{ \vec{a}_t, \vec{d}_t | &
 a_{nm}^t \leq 1-x_{nm}^{t-1}, d_{nm}^t \leq x_{nm}^{t-1}, n \in \mathcal{N}, m \in \mathcal{M} \\
& \sum\limits_{n=1}^{N} v_n (x_{nm}^{t-1} + a_{nm}^t - d_{nm}^t) \leq b_m, m \in \mathcal{M} \} \end{array},$$
where the first inequality does not allow to add a content to an SCBS cache if it is already cached in it, while the second means that a content cannot be evicted from a cache if it is not already cached in it. The third inequality does not permit the total size of the cached content in an SCBS to exceed the cache capacity.

In each stage $t = 1, ..., T-1$, the decisions regarding which contents to add and/or evict to/from the cache of the $m$th SCBS are made by solving the Bellman equation 
\begin{equation} 
\begin{split}
\begin{aligned}
&V_{t-1}(S_{t-1}) =  \min\limits_{\vec{a}_t, \vec{d}_t \in \chi_t} \Big[ I(\vec{a}_t, \vec{d}_t) +   \min\limits_{y_{nm}^u} \Big\{  \sum\limits_{n} \sum\limits_{u \in \mathcal{U}} \\
&\sum\limits_{m \in \mathcal{M}^{'}} \lambda_{n}^t c_{m}^u y_{nm}^u | \sum\limits_{j=0}^{M} y_{nm}^u = 1, y_{nm}^u \leq x_{nm}^{t} \Big\} + V_{t} (S_{t}) \Big]
\end{aligned}
\end{split}
\label{eq:bellman}
\end{equation}
where $V_t(S_t)$ represents the value of the MDP model when the system is in state $S_t$, $\lambda_n^t$ represents the number of requests in time period $t$ and $c_{m}^u$ is the content delivery cost from $m \in \mathcal{M^{'}}$ SCBS to the user $u$. The variable $y_{nm}^u$ indicates from where a content demand is satisfied. This parameter is equal to 1 when a request for content $n$ issued by a client $u$ is satisfied with the content copy cached in the $m$th SCBS, differently its value is 0. 
The parameter $I(\vec{a}_t, \vec{d}_t)$ corresponds to the penalty coming as a result of actions $\vec{a}_t$ and $\vec{d}_t$ and is defined as follows 
\begin{equation}
I(\vec{a}_t, \vec{d}_t) = \gamma \sum\limits_{n = 1}^{N} \sum\limits_{m = 1}^{M} (a_{nm}^t + d_{nm}^t),
\label{eq:penalty}
\end{equation} 
where $\gamma$ stands for the cost (per content) incurred when we add/remove a content to/from a cache. Without loss of generality, we assume that the cost to add a content to a cache is equal to the cost of evicting the content from it. Here, although we assume that the penalty is linear to the number of contents to add/evict, other models can be used to determine the penalty value. Since at the end of the horizon (e.g., at mid-night) the contents' update does not incur any penalty, two adjacent decision periods (e.g., days) can be considered independently by setting boundary condition $V_T(S_T) = 0$. Thus, during the first stage when we renew all contents free-of-charge, the penalty term $I(\vec{a}_t, \vec{d}_t)$ in (\ref{eq:bellman}) is 0 and the decision problem becomes
%
\begin{equation} 
\begin{split}
\begin{aligned}
&V_{0}(S_{0}) = \min\limits_{\vec{x}_1} \Big[ \min\limits_{y_{nm}^u} \Big\{  \sum\limits_{n} \sum\limits_{u \in \mathcal{U}} \\
& \sum\limits_{m \in \mathcal{M}^{'}} \lambda_{n}^1 c_{m}^u y_{nm}^u | \sum\limits_{j=0}^{M} y_{nm}^u = 1, y_{nm}^u \leq x_{nm}^{1} \Big\}+ V_{1} (S_{1}) \Big], %
\end{aligned}
\end{split}
\label{eq:bellman0}
\end{equation} 
\indent In our scheme, differently from the standard offline approaches where it is assumed that content requests follow a long term distribution, e.g. Zipf, we employ an auto-regressive model to estimate the expected number of content requests $\lambda_{n}^t, n \in \mathcal{N}$. For new contents, we introduce a regression model to predict their initial popularity. If we denote by $\epsilon$ a noise parameter that captures the randomness in the evolution of $\lambda_{n}^t$, the number of content requests is estimated as 
\begin{equation}
\tilde{\lambda}_{n}^t = \mu_t \cdot (\sum\limits_{\tau=1}^{H} \beta_\tau \lambda_{n}^{t-\tau}  + \epsilon),
\label{eq:lambda}
\end{equation}
where $H$ is the number of previous stages we take into account during the prediction of content requests. The coefficients $\beta_\tau, \tau = 1, ..., H-1$ correspond to the weights of each of the $H$ previous stages to the prediction process. Apparently, it is $\sum\limits_{\tau=1}^H \beta_\tau = 1$. The parameter $\mu_t$ stands for the average number of per-content requests during stage $t$; this parameter varies with the stage index to reflect the time-varying demand pattern over the day. It is worth noting that in this paper we assume that the auto-regressive model parameters, $\mu_t$, $\beta$ and the distribution for $\epsilon$ are known.   The determination of the optimal values of these parameters is an interesting problem, but it is out of the scope of this paper. This could be done considering, for example, contextual information such as language, category, publisher, etc., and this prediction could be treated as the baseline number of contents' requests per hour, $\mu_t$. Finally, we should note that in the employed auto-regressive model, the number of requests is not affected by the age of information, i.e., how long the content has been available to the content catalogue, as well as we assume that content will not disappear before stage $T$.
These are reasonable assumptions as for the results we assume a horizon of $T=24$ hours, i.e., one day. 

\section{Rolling horizon cache optimization algorithm}
\label{sec:rollinghorizon}

The online dynamic programming cache optimization model presented in the previous section suffers from the curse of dimensionality and cannot be solved exactly for realistic size problems. Specifically, at every stage $t$ an integer problem with $(M^2+2M)N$ variables should be solved. Given that the content cache status is a binary variable (e.g., a content can be cached or not in an SCBS), the discrete state space has $\Big(\sum\limits_{l=0}^{b} C_N^l\Big)^M$ components, where $b$ is the cache capacity of every SCBS (assume they all have the same size) and $C_N^l$ represents the number of combinations of $l$ elements from $N$ objects. Therefore, in order to solve the online problem using backward induction algorithm for $2$ SCBSs, $10$ contents and $b=3$, we need to enumerate the Bellman equation (\ref{eq:bellman}) into $30976$ states during every stage. To combat the dimensionality problem, we propose the \textit{rolling horizon} (RH) algorithm which is based on approximate dynamic programming (ADP) concepts and allows us to find efficient solutions. 

Our motivation lies to the fact that backward induction algorithm needs the value function (i.e., cost) for stage $t$, e.g, $V_{t}(S_{t})$, to optimize the cache decisions in stage $t-1$. If this cost is estimated for all possible states $S_{t}$, the decision problem in (\ref{eq:bellman}) only contains information for the current stage $t-1$, and thus it can be solved independently in each stage. However, the value function $V_{t}(S_{t})$ represents the future cost of following the optimal decision policy starting from state $S_{t}$, which captures both the information involvement and the optimal decision policy. Therefore, it is nearly impossible to obtain an accurate evaluation without backward recursion.

To this aim, we construct an approximation of $V_{t}(S_{t})$ using ADP ideas, which enables an efficient forward recursion. This simplification is achieved by solving explicitly an approximation of the problem over some stage horizon $\Gamma$. Both the evolution of the content popularity and the best future actions can be captured to some extent with this type of formulation. However, in order to avoid the dimensionality of the backward recursion, the fixed optimal policy obtained at stage $t$ is used for a certain number of periods determined by $\Gamma$ rather than until the terminate stage $T$. This can be viewed as an extended short-sighted approach, which considers part of the influence of the current action to the future. If the \textit{rolling horizon} algorithm is implemented for $\Gamma$ periods into the future, the decision problem at stage $t$ becomes:
\begin{equation} 
\begin{split}
\begin{aligned}
& \tilde{V}_{t-1}(S_{t-1}) \approx \min\limits_{\vec{a}_t, \vec{d}_t \in \chi_t} \Big[ I(\vec{a}_t, \vec{d}_t) + 
\min\limits_{y_{nm}^u} \Big\{  \sum\limits_{n} \sum\limits_{u \in \mathcal{U}} \\ & \sum\limits_{m \in \mathcal{M}^{'}} (\lambda_{n}^{t} + \sum\limits_{\tau = 1}^{\Gamma} \tilde{\lambda}_{n}^{t+\tau}) c_{m}^u y_{nm}^u | \sum\limits_{j=0}^{M} y_{nm}^u = 1, y_{nm}^u \leq x_{nm}^{t} \Big\}  \Big]
\end{aligned}
\end{split}
\label{eq:rollinghorizon}
\end{equation} 
where $\tilde{\lambda}_{n}^{t+\tau}$ denotes the forecasted number of requests in stage $t+\tau$ and is computed by (\ref{eq:lambda}). 

In the following subsections, we present two further simplifications of the online cache optimization problem that allow us to determine well-performing solutions with reduced complexity. For the sake of simplicity, we define the inner decision problem of (\ref{eq:rollinghorizon}) as $P$ to stand for the minimum communication cost (cost of satisfying all content demands during a period) under the current cache status. Note that without loss of generality, we replace the $(\lambda_{n}^{t} + \sum\limits_{\tau = 1}^{\Gamma} \tilde{\lambda}_{n}^{t+\tau})$ in the rolling horizon update formula (\ref{eq:rollinghorizon}) by a variable $\bar{\lambda}_{n}^{t}$ because the simplifications described in the following can be used in other ways of demand forecasting.
\begin{subequations}
\allowdisplaybreaks
\begin{align}
& P(\vec{x}^{t})  := \min  \sum\limits_{n}\sum\limits_{u \in \mathcal{U}} \sum\limits_{m \in \mathcal{M}^{'}} \bar{\lambda}_{n}^{t} c_{m}^u y_{nm}^u \\
& s.t.  \sum\limits_{m \in \mathcal{M}^{'}} y_{nm}^u = 1,  n \in \mathcal{N}, u \in \mathcal{U}   \\
&  y_{nm}^u \leq x_{nm}^{t}, y_{nm}^u \in \{0, 1\} n \in \mathcal{N}, u \in \mathcal{U}, m \in \mathcal{M}^{'}   
\end{align}
\end{subequations}

\subsection{Allow a single copy of a content to be cached in SCBSs network}
\label{sec:singlecache}
Based on the observation that there are far more contents than the total capacity of all the SCBSs, which in practice holds, we simplify the decision problem further by restricting contents to be cached at maximum in one SCBS in the network. This is an effective strategy when contents' popularity is given by a smooth distribution, i.e., the popularity of contents does not vary significantly. This simplification permits us to find near-optimal solutions with reduced cost. Under this simplification, when content $n \in \mathcal{N}$ is cached in SCBS $m\in \mathcal{M}$, all requests for it are served by the SCBS which cached the content if it is less costly than receiving it from the backhaul. Therefore, the problem $P(\vec{x}^{t})$ can be decomposed into sub-problems solved by contents. For example, for content $n \in \mathcal{N}$ we need to solve the following sub-problem:
\begin{subequations}
\begin{align}
& P_n(\vec{x}^{t}) := \min \sum\limits_{u \in \mathcal{U}} \sum\limits_{m \in \mathcal{M}^{'}} \bar{\lambda}_{n}^{t} c_{m}^u y_{nm}^u \label{eq:opt3}\\
& s.t. \sum\limits_{m \in \mathcal{M}^{'}} y_{nm}^u = 1,  u \in \mathcal{U}   \label{eq:opt31}\\
&  y_{nm}^u \leq x_{nm}^{t}, y_{nm}^u \in \{0, 1\}  u \in \mathcal{U}, m \in \mathcal{M}^{'}   \label{eq:opt32} 
\end{align}
\end{subequations}
If the content $n$ is cached in SCBS $\hat{m}$, i.e., $x_{n\hat{m}}^{t} = 1$ and each content is cached in only one SCBS, then $x_{nm}^{t} = 0, \forall m \neq \hat{m}$. This means that $y_{nm}^u = 0, \forall m \neq \hat{m}$ according to constraint \eqref{eq:opt32}. Hence, it should be $y_{n\hat{m}}^u = 1- \sum\limits_{m \neq \hat{m}} y_{nm}^u = 1$ in order to to satisfy \eqref{eq:opt31}. In this case, the minimum usage cost is $\sum\limits_{u \in \mathcal{U}} \lambda_{n}^{t} c_{\hat{m}}^u$ for each content $n$. However, if the content $n$ is not cached at any SCBSs, the minimum usage cost is trivially equal to $\sum\limits_{u \in \mathcal{U}} \lambda_{n}^{t} c_{0}^u$. Considering the above costs, we conclude that under a cache placement $x_{nm}^{t}$, the total communication cost can be computed as:
\begin{equation}
\sum\limits_{n} \lambda_{n}^{t} \sum\limits_{u \in \mathcal{U}} \left[ c_{0}^u (1-\sum\limits_{m=1}^M x_{nm}^{t}) + \sum\limits_{m=1}^M c_{m}^u x_{nm}^{t} \right]
\label{eq:closedformulacost}
\end{equation}
This removes the need to solve the problem $P$, and the minimization problem (\ref{eq:bellman}) becomes  
\begin{equation}
\begin{aligned}
\begin{split}
& V_{t-1}(S_{t-1}) = \min\limits_{\vec{a}_t, \vec{d}_t \in \chi_t} \Big\{ I(\vec{a}_t, \vec{d}_t) + \\
& \sum\limits_{n} \lambda_{n}^{t} \sum\limits_{u \in \mathcal{U}} \Big[ c_{0}^u + \sum\limits_{m=1}^M (c_{m}^u-c_{0}^u) x_{nm}^{t} \Big] + V_{t} (S_{t}) \Big\}.
\label{eq:Bellman2}
\end{split}
\end{aligned}
\end{equation}
\indent Note that the simplified problem in (\ref{eq:Bellman2}) does not contain content delivery decision variables $y$, and hence the size of the action space is greatly reduced from $(M^2+2M)N$ to $2MN$. If we further consider the simplification incurred by the rolling-horizon approximation described by (\ref{eq:rollinghorizon}), the cache and delivery optimization problem can be written as
{
\begin{subequations}
\allowdisplaybreaks
\begin{align}
& \min  \sum\limits_{n \in \mathcal{N}} \sum\limits_{m \in \mathcal{M}} I(\vec{a}_t, \vec{d}_t) + \sum\limits_{n} (\lambda_{n}^t + \sum\limits_{\tau = 1}^{\Gamma} \tilde{\lambda}_{n}^{t+\tau}) \nonumber \cdot\\
&\cdot \sum\limits_{u \in \mathcal{U}} \left[ c_{0}^u + \sum\limits_{m=1}^M (c_{m}^u-c_{0}^u) (x_{nm}^{t-1} + a_{nm}^t - d_{nm}^t) \right] \label{eq:opt5}\\
s.t. & \;a_{nm}^t \leq 1-x_{nm}^{t-1}, \forall n \in \mathcal{N}, \forall m \in \mathcal{M} \label{eq:opt52}\\
& \;d_{nm}^t \leq x_{nm}^{t-1}, \forall n \in \mathcal{N}, \forall m \in \mathcal{M} \label{eq:opt53}\\
& \;\sum\limits_{n=1}^{N} v_n (x_{nm}^{t-1} + a_{nm}^t - d_{nm}^t) \leq b_m, \forall m \in \mathcal{M} \label{eq:opt54}\\  
& \;\sum\limits_{m \in \mathcal{M}} (x_{nm}^{t-1} + a_{nm}^t - d_{nm}^t) \leq 1, \forall n \in \mathcal{N}  \label{eq:opt55}\\
& \;a_{nm}^t, d_{nm}^t \in \{0, 1\}, \forall n \in \mathcal{N}, \forall m \in \mathcal{M} \label{eq:opt56}
\end{align}
\end{subequations}
}
We can see that all the constraints of the original online minimization problem (\ref{eq:bellman}) are also constraints of the simplified problem in (\ref{eq:opt5}), but in addition we have constraint \eqref{eq:opt55} which prohibits the caching of a content in two or more SCBSs. This constraint enables the use of the closed-form formula (\ref{eq:closedformulacost}) for computing the optimal usage cost, and thus decreases the cost of determining the optimal solution. As we will see in the evaluation section in the majority of the cases the proposed simplification results in the same cache allocation with that of the original problem (\ref{eq:bellman}) under the assumption that content popularity follows a smooth distribution. 

\subsection{Limit the number of allowed content replacements}
\label{sec:singlecacheHeuristic}

Despite the simplification considered in the previous subsection, the action space could be still large in size. To overcome this problem, we propose a greedy heuristic summarized in Algorithm~\ref{alg:replace} to solve (\ref{eq:rollinghorizon}), which examines contents with decreasing order of popularity. $\Delta(x_{j \leftarrow i,k}^t)$ corresponds to the difference in the cost when content $j$ is replaced by content $i$ at the cache of the $k$th SCBS. A content replacement happens only if the replacements of a content in an SCBS leads to the largest reduction of total usage cost, regardless of whether contents involved in this replacement are cached anywhere else in the network. This means multiple SCBSs may have a copy of the same content, which is different from the assumption used in Section \ref{sec:singlecache}. The cost reduction is calculated by considering a fixed length stage horizon into the future, as what we have used in (\ref{eq:rollinghorizon}). As each potential replacement involves only two contents, i.e. the added and the evicted one, the penalty of cache update and the potential cost reduction can be evaluated very quickly.

\begin{algorithm}[t]
\baselineskip=10pt
\caption{\textbf{Greedy heuristic for content replacement}} 
\label{alg:replace}
\begin{algorithmic}[1]
\STATE Input: maximum number of content replacements per cache $r$ 
\STATE Sort contents in an decreasing order of the forecasted number of downloads
\FOR {$i = 1,\ldots,r$}
\FOR {$k = 1,\ldots,M$}
\STATE  $$\Delta(a_{ik}^t) = \gamma + [P_i(x_{ik}^t=1) - P_i(x_{ik}^t=0)]$$ 
\FOR {each content $j$ that is cached on SCBS $k$ ($x_{jk}^{t-1} = 1$)}
\STATE $$\Delta(d_{jk}^t) = \gamma + [P_j(x_{jk}^t=0) - P_j(x_{jk}^t=1)]$$ 
\STATE {$$\Delta(x_{j \leftarrow i,k}^t) = \Delta(a_{ik}^t) + \Delta(d_{jk}^t)$$}
\ENDFOR
\ENDFOR
\IF {($\min \Delta(x_{j \leftarrow i,k}^t)<0$)}
\STATE Carry out the corresponding replacement.  
\ENDIF
\ENDFOR
\end{algorithmic}
\end{algorithm}

As shown in Algorithm~\ref{alg:replace}, the greedy heuristic turns the optimization problem into a sequential decision-making problem by content. This is possible because demands for contents are independent of each other. The inner optimization $P_n$ for each content $n$ calculates the cost of satisfying customer demands given the existing caching locations. The outer problem, which selects the optimal $\vec{a}_t, \vec{d}_t$ values, is subject to the capacity constraint of every SCBS and cannot be decomposed. Nevertheless, assuming that content updates are done via replacement, the capacity constraint is always satisfied. Therefore, the overall optimization problem boils down to a content-by-content replacement problem where more popular contents replace the less popular ones in each SCBS. Please note that when $P_n$ problems are solved (lines 5 and 7 in Algorithm~\ref{alg:replace}), the popularity is forecasted as $\bar{\lambda}_n^t = (\lambda_{n}^t + \sum\limits_{\tau = 1}^{\Gamma} \tilde{\lambda}_{n}^{t+\tau})$.

The greedy heuristic proposed in Algorithm~\ref{alg:replace} is of low computational complexity, as it only requires to compare the parts of the value function which are relevant to the added and evicted contents in an SCBS. Therefore, the computational complexity of the algorithm is linear with respect to the number of allowed replacements $r$ per decision stage. As the greedy algorithm calculates the cost reductions that are coming from each replacement decision separately taking into account content popularity forecasting, all contents can be potentially replaced per decision stage. This type of heuristic is derived naturally by examining the structure of the optimization problem in (\ref{eq:rollinghorizon}) and has great potential of providing very good solutions given a large enough number of cache updates is allowed.  

\section{Experimental results}
\label{sec:results}

In this section, we evaluate the performance of the proposed \textit{rolling horizon} algorithm presented in Section \ref{sec:rollinghorizon}. We compare the \textit{rolling horizon} algorithm with a scheme that decides every $T$ stages the cache placement considering a Zipfian distribution and then update it based on the LRU policy and with two other online cache updating policies, i.e., a myopic policy and an one-step improvement policy, which are described in Section \ref{sec:benchmark}. 


\subsection{Experimental results settings}

We evaluate the performance of all schemes under comparison for various settings summarized in Table \ref{tab:networksetting}, where the ``Ratio'' column corresponds to the ratio between the cumulative size of the contents in the content catalogue and the total cache capacity of the SCBSs. The third column in Table \ref{tab:networksetting} depicts the SCBS network topology. For all cases, we consider that the SCBSs form a grid network where a crossover point indicates the existence of an SCBS. In addition, we assume that all the SCBSs are connected with a direct link to the MCBS, which for the sake of simplicity is not depicted in Table \ref{tab:networksetting}. The communication cost to deliver a demanded content to a client depends on the distance between the SCBS at where a client is located, and the base station (either SCBS or MCBS) from where the requested content is obtained. Specifically, the communication cost is linearly dependent with respect to the length of the shortest path between them, while the downloading cost of a content from the MCBS $c_{m0}^u$ is set to 20.  

{\setstretch{0.8}
\begin{table*}[t]
\caption{Evaluation scenarios settings}
\begin{center}
\small
\scriptsize
\begin{tabular}{lccccccc}
\hline
\hline
Index & \#SCBS & Network topology & \#Contents & Penalty($\gamma$) & Capacity & Ratio & \#New contents per hour\\
\hline
\textit{Ins 1.1}		& & 	 & 	 	& 		& 1		& 30\%	& \\
\textit{Ins 1.2}			& 3 & $\left[ {\begin{array}{ccc}
   \bullet & \bullet & \bullet\\
  \end{array} } \right]$
	 & 10 	& 100 & 2		& 60\%	& 1\\
\textit{Ins 1.3}			& &	 & 	 	&		& 3		& 90\%	& \\
\textit{Ins 1.4}			& &	 & 	 	&		& 4		& 120\%	& \\
\hline
\textit{Ins 2.1}		& & 	 & 	 	& 		& 10		& 30\%	& \\
\textit{Ins 2.2}			& 3 & $\left[ {\begin{array}{ccc}
   \bullet & \bullet & \bullet\\
  \end{array} } \right]$
	 & 100 	& 100 & 20		& 60\%	& 2\\
\textit{Ins 2.3}			& &	 & 	 	&		& 30		& 90\%	& \\
\textit{Ins 2.4}			& &	 & 	 	&		& 40		& 120\%	& \\
\hline
\textit{Ins 3.1} 		& &	 & 	 	&		& 4		& 24\%	& \\
\textit{Ins 3.2} 		& &	 & 	 &		& 8		& 48\%	& \\
\textit{Ins 3.3} 		& 6 & $\left[ {\begin{array}{ccc}
   \bullet & \bullet & \bullet\\
	\bullet & \bullet & \bullet\\
  \end{array} } \right]$ & 100	&	100 & 12	  & 72\%	&	2\\
\textit{Ins 3.4} 		& &	 & 	 &			& 17	& 102\%	& \\
\hline
\textit{Ins 4.1} 		& &	 & 	 	&		& 20		& 24\%	& \\
\textit{Ins 4.2} 		& &	 & 	 &		& 40		& 48\%	& \\
\textit{Ins 4.3} 		& 6 & $\left[ {\begin{array}{ccc}
   \bullet & \bullet & \bullet\\
	\bullet & \bullet & \bullet\\
  \end{array} } \right]$ & 500	&	100 & 60	  & 72\%	&	5\\
\textit{Ins 4.4} 		& &	 & 	 &			& 80	& 96\%	& \\
\hline
\textit{Ins 5.1}			& &	 & 	 &			& 10	& 24\%	& \\
\textit{Ins 5.2}			& &	 & 	 &			& 20	& 48\%	& \\
\textit{Ins 5.3}			& 12 & $\left[ {\begin{array}{cccc}
   \bullet & \bullet & \bullet & \bullet\\
	\bullet & \bullet & \bullet & \bullet\\
	\bullet & \bullet & \bullet & \bullet\\
  \end{array} } \right]$ & 500	&	100 & 30	& 72\%	& 5\\
\textit{Ins 5.4}			& &	 & 	 &			& 40	& 96\%	& \\
\hline
\textit{Ins 6.1}			& &	 & 	 &			& 20	& 24\%	& \\
\textit{Ins 6.2}			& &	 & 	 &			& 40	& 48\%	& \\
\textit{Ins 6.3}			& 12 & $\left[ {\begin{array}{cccc}
   \bullet & \bullet & \bullet & \bullet\\
	\bullet & \bullet & \bullet & \bullet\\
	\bullet & \bullet & \bullet & \bullet\\
  \end{array} } \right]$ & 1000	&	100 & 60	& 72\%	& 10\\
\textit{Ins 6.4}			& &	 & 	 &			& 80	& 96\%	& \\
\hline
\textit{Ins 7.1} 		& &	 & 	 &			& 17	& 25.5\% & \\ 
\textit{Ins 7.2} 		& &	 & 	 &			& 33	& 49.5\% & \\ 
\textit{Ins 7.3} 		& 15 & $\left[ {\begin{array}{ccccc}
   \bullet & \bullet & \bullet & \bullet & \bullet\\
	\bullet & \bullet & \bullet & \bullet & \bullet\\
	\bullet & \bullet & \bullet & \bullet & \bullet\\
  \end{array} } \right]$ & 1000	&	100 & 50	& 75\% & 10\\ 
\textit{Ins 7.4} 		& &	 & 	 &			& 66	& 99\% & \\ 
\hline
\hline
\end{tabular}
\label{tab:networksetting}
\end{center}
\end{table*}
}
%
The parameter $\gamma$ in Table \ref{tab:networksetting} is used to compute the cost (penalty) one has to pay for each content update in the cache, i.e., when a cached content in an SCBS is replaced by another and is calculated by (\ref{eq:penalty}). This parameter is associated with the potential delay of meeting users' requests for adding and evicting the contents. This cost should be larger than the cost of downloading a content directly from the MCBS, which is the minimum delay (cost) one has to wait for updating a content in an SCBS. Also, this cost should be smaller than the cost incurred when all contents' requests are served by the MCBS in one stage, as otherwise there would not be any reason to alter the current cache allocation. Thus, for all network settings and content update ratios, we set the parameter $\gamma$ to be equal to 100, which is five times the cost of requesting a content from the MCBS. For all experiments, the $\beta_\tau$ parameters used from our forecasting model in (\ref{eq:lambda}) are $(0.6, 0.3, 0.1)$. Finally, we consider hourly cache updates and $T=24$.

\subsection{Comparison schemes}
\label{sec:benchmark}

Prior to presenting the experimental results, we introduce the comparison schemes. The first comparison scheme decides which contents to cache in the SCBSs in two phases: an offline phase where the off-peak network caching update policy is made assuming that content popularity follows a Zipfian distribution, and an online cache updating phase based on LRU policy. For both cache update phases, the content placement in the SCBSs' caches is decided centrally considering the SCBSs' communication model described in Section \ref{sec:net_setting}. We also compare with several alternative approximations of the proposed dynamic programming model (\ref{eq:bellman}). These policies aim at incorporating future information through an approximated value-to-go function, i.e. the $V_{t}(S_{t})$ in (\ref{eq:bellman}). For these policies, the cache update decisions are made by solving (\ref{eq:bellman}) with an approximated $V_{t}(S_{t})$, which is policy-specific, and, therefore, the simplifications discussed in Section \ref{sec:singlecache} and \ref{sec:singlecacheHeuristic} are also applicable to them. 

\subsubsection{Offline cache optimization with online updates based on LRU}
\label{offline}


During the offline phase cache update decisions are made assuming that content popularity follows a Zipf distribution with a known skewness parameter. Let $p_n$ represent the popularity of content $n$, the optimization problem is formally expressed as
\begin{subequations}
\allowdisplaybreaks
\begin{align}
\min & \sum\limits_{n \in \mathcal{N}} p_n \Big(\sum\limits_{u \in \mathcal{U}} \sum\limits_{m \in \mathcal{M}^{'}} c_{m}^u y_{nm}^u\Big) \label{eq:opt1}\\
s.t. & \sum\limits_{n \in \mathcal{N}} v_{n} x_{nm} \leq b_{m},  m \in \mathcal{M} \label{eq:opt11}\\
& \sum\limits_{m \in \mathcal{M}^{'}} y_{nm}^u = 1,  u \in \mathcal{U}, n \in \mathcal{N}  \label{eq:opt12}\\
&  y_{nm}^u \leq x_{nm},  u \in \mathcal{U}, m \in \mathcal{M}^{'}, n \in \mathcal{N}  \label{eq:opt13} \\ 
& x_{nm} \in \{0, 1\},  n \in \mathcal{N}, m \in \mathcal{M} \\
& y_{nm}^u \in \{0, 1\},  u \in \mathcal{U}, m \in \mathcal{M}^{'}, n \in \mathcal{N} 
\end{align}
\end{subequations}
where the constraint \eqref{eq:opt11} corresponds to the capacity constraint, which prevents the total volume of contents cached in a SCBS from exceeding its capacity. The constraint \eqref{eq:opt12} ensures that all customer requests are satisfied. Finally, the constraint \eqref{eq:opt13} imposes that a content can be obtained from a SCBS only if it is cached in it. 


In the online phase, the cached content at the SCBSs is updated by applying LRU policy. 
The online cache optimization aims at capturing sudden content popularity changes and demands for contents that were not previously available in the content catalogue, so as to minimize the backhaul link load. Hence, in every decision stage the least popular cached content is replaced by the most popular one as observed in this stage. As in each stage, there might be new contents or contents that their popularity changed significantly, multiple content replacements are allowed. Depending on whether caching multiple copies of the same content is allowed within the network, we consider two variations of the LRU policy: $LRU(S)_r$ and $LRU(M)_r$, where $r$ stands for the number of allowed replacements in each SCBS. $S$ and $M$ indicate whether a single copy or multiple copies per content are allowed, respectively. We should note that the determination of the optimal $r$ value is out the scope of this paper. 

\subsubsection{Myopic policy}

The \textit{Myopic} policy is a short-sighted approach which only considers the immediate cost when solving the Bellman equation (\ref{eq:bellman}). This means that the future expected cost, $V_{t} (S_{t})$, by carrying out the optimal policy in all future stages, is set to zero. Therefore, the cache updates are decided by solving the following problem:
\begin{equation} 
\begin{aligned}
\begin{split}
& \hat{V}_{t-1}(S_{t-1}) \approx \min\limits_{\vec{a}_t, \vec{d}_t \in \chi_t} \Big[ I(\vec{a}_t, \vec{d}_t) + \min\limits_{y_{nm}^u} \Big\{  \sum\limits_{n} \sum\limits_{u \in \mathcal{U}} \\ 
&\sum\limits_{m \in \mathcal{M}^{'}} \lambda_{n}^t c_{m}^u y_{nm}^u | \sum\limits_{m=0}^{M} y_{nm}^u = 1, y_{nm}^u \leq x_{nm}^{t} \Big\} \Big].
\end{split}
\end{aligned}
\end{equation}

\subsubsection{One-step improvement policy}

Another policy we use for evaluation purpose is the  \textit{One-step} improvement policy. This policy works similarly to the \textit{Myopic} policy in that it solves the Bellman equation for every SCBS independently and then decides the contents to update based on the computed cost values. However, differently from the \textit{Myopic} policy, it considers the long-run usage cost in addition to the immediate cost. For the computation of the long-run usage cost, it is assumed that both the content popularity and the caching plan will not be updated in the future. Under this assumption, the long-run usage cost can be captured by a Zipfian distribution, which estimates the average popularity of every content over the entire decision horizon based on the encountered content requests. In order to capture the evolution of the number of requests per stage, the Zipfian distribution is rescaled. Hence, if the total number of content requests is constant over the future periods and a percentage $p_n$ of them is for content $n$, the expected number of requests for content $n$ in the future periods evolves as follows:
$$\tilde{\lambda}_{n}^{t+\tau} = p_{n} (\sum\limits_{\hat{n}=1}^N \lambda_{\hat{n}}^t), \tau = 1, ..., T-t,$$
Then, the cache updates are decided by solving 
\begin{equation} 
\begin{aligned}
\begin{split}
&\bar{V}_{t-1}(S_{t-1}) \approx \min\limits_{\vec{a}_t, \vec{d}_t \in \chi_t} \Big[ I(\vec{a}_t, \vec{d}_t) + \min\limits_{y_{nm}^u} \Big\{ \sum\limits_{n} \sum\limits_{u \in \mathcal{U}} \\
&\sum\limits_{m \in \mathcal{M}^{'}} \lambda_{n}^t c_{m}^u y_{nm}^u | \sum\limits_{m=0}^{M} y_{nm}^u = 1, y_{nm}^u \leq x_{nm}^{t} \Big\} + \min\limits_{y_{nm}^u} \Big\{ \sum\limits_{n} \sum\limits_{u \in \mathcal{U}}\\
& \sum\limits_{m \in \mathcal{M}^{'}} \sum\limits_{\tau=1}^{T-t} \tilde{\lambda}_{n}^{t+\tau} c_{m}^u y_{nm}^u | \sum\limits_{j=0}^{M} y_{nm}^u = 1, y_{nm}^u \leq x_{nm}^{t} \Big\} \Big] 
\label{eq:onestep}
\end{split}
\end{aligned}
\end{equation}

By observing (\ref{eq:onestep}), we note that the inner decision problem (selection of optimal $y$ values) is split into two subproblems: one aiming to optimize the immediate cost (second summand  in (\ref{eq:onestep})) and another aiming to minimize the approximation of the cost-to-go, $V_{t} (S_{t})$ (third summand in (\ref{eq:onestep})). The two subproblems can be merged as follows:
%
\begin{equation} 
\begin{aligned}
\begin{split}
& \bar{V}_{t-1}(S_{t-1}) \approx \min\limits_{\vec{a}_t, \vec{d}_t \in \chi_t} \Big[ I(\vec{a}_t, \vec{d}_t) + \min\limits_{y_{nm}^u} \Big\{ \sum\limits_{n} \sum\limits_{u \in \mathcal{U}} \\
& \sum\limits_{m \in \mathcal{M}^{'}} (\lambda_{n}^t + \sum\limits_{\tau=1}^{T-t} \tilde{\lambda}_{n}^{t+\tau}) c_{m}^u y_{nm}^u | \sum\limits_{m=0}^{M} y_{nm}^u = 1, y_{nm}^u \leq x_{nm}^{t} \Big\} \Big] 
\end{split}
\end{aligned}
\end{equation} 

\subsection{Numerical Results}

In this section, we examine the performance of the proposed online cache optimization schemes described in Sections \ref{sec:onlinemodel} and \ref{sec:rollinghorizon} and
compare them with those presented in Section \ref{sec:benchmark}. All the reported results are averages of 100 simulations, i.e., realizations of different content requests. 

\subsubsection{Performance evaluation for exact solution of Bellman equation}

We first examine the performance of the online cache optimization policies (i.e., the generic dynamic programming framework and the \textit{rolling horizon} policies RH1-RH3, where the number corresponds to the size of the horizon) for updating the cached contents in the SCBSs presented in Section \ref{sec:rollinghorizon} and compare them with the methods discussed in Section \ref{sec:benchmark}. 
We should note that despite the fact that the \textit{Myopic} and the \textit{One-step improvement} policies approximate differently the value-to-go function $V_{t}(S_{t})$ in (\ref{eq:bellman}), for the sake of deriving conclusions that are not affected by the solution approach, we assume that the best action of (\ref{eq:bellman}) with an underlying policy is found by solving optimally the optimization problem without any simplifications. As the optimal actions can be computed only for problems of very small size due to the computational complexity of solving the integer program, we present results only for the settings \textit{Ins 1.$k$}, \textit{Ins 2.$k$}, and \textit{Ins 3.$k$},  $k \in \{1,2,3,4\}$(see Table \ref{tab:networksetting}). 

{\setstretch{0.8}
\begin{table*}[t]
\caption{Usage cost for the online models assuming exact solution of (\ref{eq:bellman})}
\begin{center}
\small
\scriptsize
\begin{tabular}{l|c|cc|ccccc|c}
\hline
\hline
  & & \multicolumn{2}{|c|}{\centering  \textbf{Offline(Zipf) + LRU }} & \multicolumn{5}{|c|}{\centering  \textbf{Dynamic programming }} & \textbf{Offline}\\
\hline
Index 			& LB 		& $LRU(S)_r$ & $LRU(M)_r$	& Myopic 	& One-step 	& RH1 & RH2 & RH3 & $x_0$\\
\hline
\textit{Ins 1.1} 		& 0.00	& 0.6983 & 2.2488 & 0.2824 & 0.6817 & \underline{0.2559} & 0.2727 & 0.3013 & 1.00 \\
\textit{Ins 1.2} 		& 0.00	& 0.4989 & 2.2919 & \underline{0.1787} & 0.4132 & 0.1812 & 0.2093 & 0.2224 & 1.00 \\
\textit{Ins 1.3} 		& 0.00  & 0.2346 & 1.9823 & 0.1241 & 0.1745 & \underline{0.1084} & 0.1179 & 0.1285 & 1.00 \\
\textit{Ins 1.4} 		& 0.00 	& \underline{0.1027} & 2.4033 & 0.1338 & 0.1446 & 0.1052 & 0.1055 & 0.1067  & 1.00 \\ 
\hline
\textit{Ins 2.1} 		& 0.00	& 0.9045 & 3.5173 & 0.4392 & 0.7614 & \underline{0.3450} & 0.3667 & 0.4054 & 1.00 \\ 
\textit{Ins 2.2} 		& 0.00	& 0.8026 & 4.0877 & 0.3454 & 0.4677 & \underline{0.2804} & 0.2977 & 0.3292 & 1.00 \\ 
\textit{Ins 2.3} 		& 0.00 	& 0.4957 & 4.4549 & 0.1594 & 0.2945 & \underline{0.1564} & 0.1774 & 0.1995 & 1.00 \\ 
\textit{Ins 2.4} 		& 0.00 	& 0.1041 & 5.3883 & 0.1350 & 0.2170 & 0.0997 & 0.0982 & \underline{0.0976} & 1.00 \\ 
\hline
\textit{Ins 3.1} 		& 0.00	& 0.9017 & 3.6483 & 0.4489 & 0.8109 & \underline{0.3403} & 0.3590 & 0.3954 & 1.00 \\ 
\textit{Ins 3.2} 		& 0.00	& 0.9117 & 4.4121 & 0.3848 & 0.6879 & \underline{0.2942} & 0.3090 & 0.3437 & 1.00 \\ 
\textit{Ins 3.3} 		& 0.00 	& 0.7353 & 4.6935 & 0.2733 & 0.4700 & \underline{0.2326} & 0.2496 & 0.2786 & 1.00 \\ 
\textit{Ins 3.4} 		& 0.00 	& 0.1547 & 5.2946 & 0.1367 & 0.2141 & 0.0974 & 0.0912 & \underline{0.0904} & 1.00 \\ 
\hline
\hline
\end{tabular}
\label{tb:resultsOnline}
\end{center}
\end{table*}
}


{\setstretch{0.8}
\begin{table*}[t]
\caption{Cache Hit Ratio for online models assuming exact solution of (\ref{eq:bellman})}
\begin{center}
\small
\scriptsize
\begin{tabular}{l|c|cc|ccccc|c}
\hline
\hline
  & & \multicolumn{2}{|c|}{\centering  \textbf{Offline(Zipf) + LRU }} & \multicolumn{5}{|c|}{\centering  \textbf{Dynamic programming }} & \textbf{Offline}\\
\hline
Index 			& LB 		& $LRU(S)_r$ & $LRU(M)_r$	& Myopic 	& One-step 	& RH1 & RH2 & RH3 & $x_0$\\
\hline
\textit{Ins 1.1} 		& 0.0853	& 0.0883 & \underline{0.0970} & 0.0794 & 0.0850 & 0.0829 & 0.0819	& 0.0861 & 0.0652 \\ 
\textit{Ins 1.2} 		& 0.1506	& 0.1546 & \underline{0.1769} & 0.1431 & 0.1460 & 0.1462 & 0.1497	& 0.1500 & 0.0944 \\ 
\textit{Ins 1.3} 		& 0.1748 	& 0.2112 & \underline{0.2540} & 0.1872 & 0.1925 & 0.1959 & 0.1989 & 0.1970 & 0.0775 \\ 
\textit{Ins 1.4} 		& 0.2792 	& 0.2451 & 0.2828 & 0.2303 & \underline{0.2831} & 0.2498 & 0.2626 & 0.2696 & 0.1346\\ 
\hline
\textit{Ins 2.1} 		& 0.0895 	& 0.0980 & \underline{0.1147} & 0.0870 & 0.0926 & 0.0922 & 0.0937 & 0.0947 & 0.0703\\ 
\textit{Ins 2.2} 		& 0.1519  & 0.1667 & \underline{0.2078} & 0.1466 & 0.1533 & 0.1536 & 0.1567 & 0.1568 & 0.1284 \\ 
\textit{Ins 2.3} 		& 0.1877  & 0.2039 & \underline{0.2893} & 0.1938 & 0.1928 & 0.1971 & 0.1971 & 0.1991 & 0.1481 \\ 
\textit{Ins 2.4} 		& 0.2819  & 0.2639 & \underline{0.3596} & 0.2692 & 0.2963 & 0.2725 & 0.2791 & 0.2782 & 0.2230\\ 
\hline
\textit{Ins 3.1} 		& 0.0617	& 0.0617 & \underline{0.0750} & 0.0561 & 0.0617 & 0.0595 & 0.0593 & 0.0610	& 0.0430 \\ 
\textit{Ins 3.2} 		& 0.1071	& 0.1073 & \underline{0.1391} & 0.0981 & 0.1074 & 0.1022 & 0.1027 & 0.1045	& 0.0837 \\ 
\textit{Ins 3.3} 		& 0.1491  & 0.1438 & \underline{0.1980} & 0.1306 & 0.1431 & 0.1315 & 0.1370 & 0.1374 & 0.1155 \\ 
\textit{Ins 3.4} 		& 0.1679 	& 0.1652 & \underline{0.2616} & 0.1710 & 0.1758 & 0.1322 & 0.1737 & 0.1384 & 0.1275\\ 
\hline
\hline
\end{tabular}
\label{tb:CHR_online}
\end{center}
\end{table*}
}
%
Table \ref{tb:resultsOnline} shows the proportional performance (usage cost) of an evaluated policy compared to the lower bound $LB$ and the offline policy as described in Section \ref{offline}. Specifically, the proportional performance of an evaluated policy is calculated as $(* - LB)/(x_0 - LB)$, where $x_0$ indicates the cost of using the optimal offline cache decisions for the entire horizon without allowing any online updates. $LB$ is equivalent to the cost of a scheme where accurate demands are known before updating the online caching and 
cache updates are done in all the stages before the end of the horizon $T$ without introducing any additional cost, i.e., the penalty cost $\gamma$ is set to 0. Hence, $LB$ can be seen as an unreachable lower bound of any practical policy. For the $LRU(S)_r$ and $LRU(M)_r$ policies, we allow potentially all the contents to be updated in each decision stage, which means that $r$ is equal to the SCBS's cache size. However, we should note that the actual number of content replacements depends on the content popularity.

From the usage cost results presented in Table \ref{tb:resultsOnline}, we can see that apart from $LRU(M)_r$, all other online schemes achieve a usage cost lower than $x_0$ for all testing scenarios. This justifies the benefits coming from an online update of the cached contents. In most testing scenarios, the proposed \textit{rolling horizon} method (RH1, RH2, RH3) outperforms $LRU(S)_r$ and $LRU(M)_r$ policies significantly. From the proposed online schemes, the best performance is achieved by RH1, which assumes a horizon of one. This is attributed to the fact that the longer we perform the forecasting, the larger is the error introduced because of the imperfectness of forecasting. 


Further, from Table \ref{tb:resultsOnline} we can observe that the performance gap between RH policies and \textit{Myopic} and $LRU(S)_r$ policy becomes smaller as SCBSs cache capacity increases. This is because when SBCSs capacity is large enough, all contents' requests can be accommodated by the caches of local SCBSs. Thus, for larger capacity values, content popularity evolution becomes less important, as there is sufficient space to cache most of or even all the contents. In such a case, close to optimal content update decisions can be made using only the immediate information. Differently, when capacity is very limited, the consideration of looking at several stages ahead leads to large usage cost reduction. Besides, from the results is obvious that in most cases the $LRU(S)_r$ policy, without considering the explicit optimization model (\ref{eq:bellman}), performs worse than the \textit{Myopic} policy, which solves the optimization model ignoring the future usage cost. This happens because in contrast to  $LRU(S)_r$ policy which bases its decisions only on recent demands, the \textit{Myopic} policy decides caching and delivery by optimizing (\ref{eq:bellman}). It is also worth to note that although we would expect \textit{One-step} improvement policy to perform better than the \textit{Myopic}, as it considers the immediate usage cost together with an estimation of the future cost, this does not happen because \textit{One-step} improvement policy assumes a Zipfian model to capture the future usage cost and no future updates of the cached contents, which leads to inaccurate approximation of the future cost $V_t(S_{t})$. 

Next, we examine the cache hit ratio achieved by all the schemes under comparison and the results are shown in Table \ref{tb:CHR_online}. We can observe that $LRU(M)_r$ achieves the highest or very close to the highest cache hit ratio although it performs worse than the other schemes in terms of cost, as shown in Table \ref{tb:resultsOnline}. This increased cache hit ratio does not lead to reduced cost even when cumulative SCBS cache capacity is higher than the total size of contents in which case multiple copies of contents should exist in the network. This is because, in a collaborative network, the policy with the minimal usage cost does not necessarily coincide with the one achieving the highest cache hit ratio. Serving more content requests from SCBSs close to the users require multiple copies, which contradicts with the aim of collaborative caching to maximize the ``re-use'' of the cached content (i.e., serve users with content cached in other SCBSs). 

\subsubsection{Performance evaluation of allowing a single copy of every content}
\label{sec:oneCopy}

In the previous subsection, we have seen that the proposed \textit{rolling horizon} policies outperform all the other schemes in terms of usage cost. Here, we investigate the impact on the performance of the approximation policy proposed in Section \ref{sec:singlecache} where we allow a single copy of a content in the SCBSs network. Considering this simplification, we minimize the usage cost by solving (\ref{eq:Bellman2}) instead of (\ref{eq:bellman}). This approximation model removes the cache delivery $y$ variables from the decision problem, and, therefore, largely reduces the problem size and the solution time.

\begin{figure*}[t]
	\centerline{\includegraphics[width=0.95\textwidth]{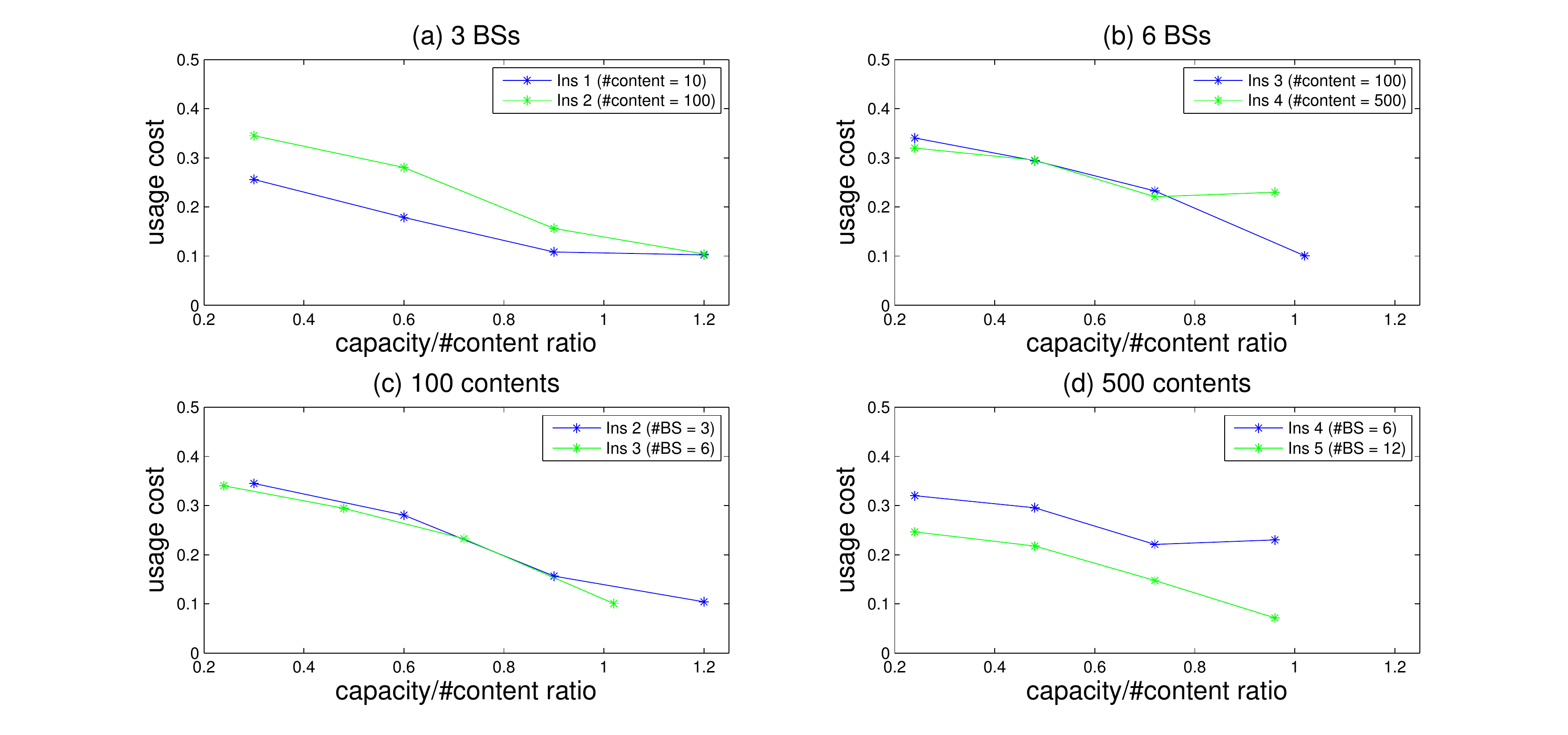}}
	\caption{Numerical results for online model when approximate solution is obtained by solving (\ref{eq:Bellman2}).}
  \label{fig:results_online_removey}
\end{figure*}

The evaluation results are summarized in Fig. \ref{fig:results_online_removey}. 
By comparing these results with the ones in Table \ref{tb:resultsOnline}, we observe that the results are nearly the same with that of solving the decision problem optimally given that the benchmark policies, say $x_0$ and $LB$, do not change with the reformulation. However, the solution time is largely reduced from 5 hours to 7 minutes, for performing 100 simulations of \textit{Ins 3} settings. This method needs in average 0.18 seconds to find the optimal cache update policy for \textit{Ins 3} in every stage. 

Interestingly, the performance of the \textit{One-step} policy improves when contents are restricted to be cached in only one SCBS in the network. This is explained by the fact that the longer the prediction period it considers, the larger are the differences in the total demands of contents, which results in some contents' popularity to be overestimated. Differences in the usage cost reduction can also be noted when cache capacity is large. This happens as in this case optimally the content should have been cached in multiple SCBSs and because of the restriction of one-copy per content restriction, an error is introduced in the reformulation. To overcome this problem, in the next section we test the heuristic replacement approach, which facilitates a fast solution for the online decision problem without imposing the one-copy per-content condition. 

\begin{figure*}[t]
\centering
\includegraphics[width=0.95\textwidth]{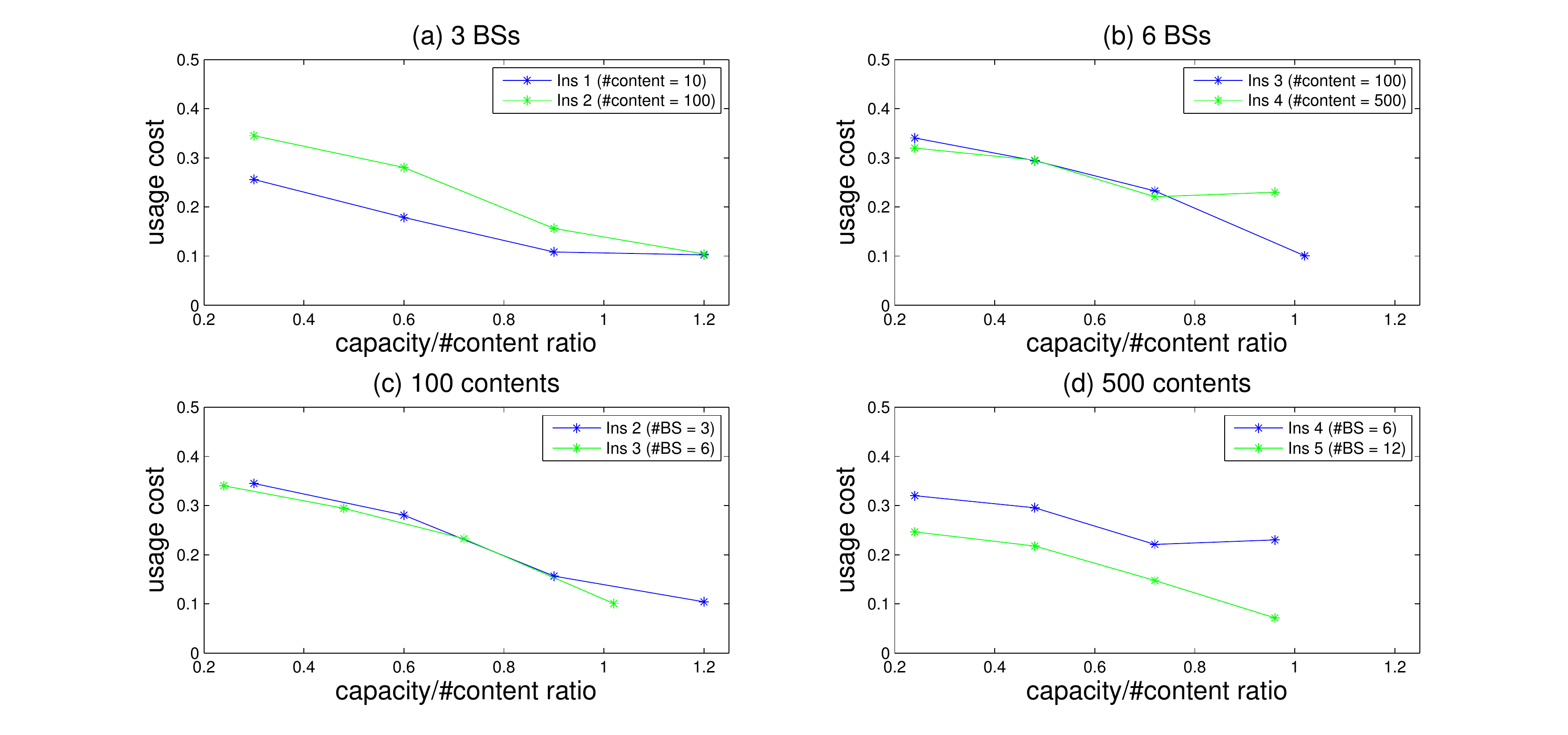}
\caption{Comparison of usage cost over different parameter settings for online models, when approximate solution is obtained by solving (\ref{eq:Bellman2}).}
\label{fig:online_compare1}
\end{figure*}

Next, we investigate the impact of the network topology and the content catalogue size to the usage cost. First, we examine the effect of the content catalogue size. Figs. \ref{fig:online_compare1}(a) and (b), which correspond to network topologies with three and six SCBSs, respectively, show the proportional usage cost of the best performing \textit{rolling horizon} policy out of all tested. From these figures, we can see that for any capacity/content ratio a smaller catalogue size leads to improved usage cost for the RH policies compared to the cost of the offline caching. Second, we examine the effect of various topologies for fixed content catalogue sizes. The results are depicted in Figs. \ref{fig:online_compare1}(c) and (d), from where we can see that the higher is the number of SCBSs, the better is the performance of the RH policy. This can be explained as follows: the larger the network topology is, the higher delivery cost it incurs, and the higher penalty one receives by sub-optimal content cache placements and updates. Therefore, larger profits are expected by updating the cached content in the SCBSs by considering both the immediate cost and a future prediction of it.  


\subsubsection{Performance evaluation of limiting the number of content replacements}

In this section, we evaluate the performance of the solution proposed in Section \ref{sec:singlecacheHeuristic}. 
This heuristic solution solves the cache optimization problem by setting the maximum number of content replacements to be equal to the total cache capacity of the network (cumulative SCBSs cache capacity). 
Therefore, potentially all the contents that are currently cached in any SCBS (including all its copies in different SCBSs) can be updated at every stage. This updating scheme is used indifferently on all dynamic programming policies, however, the forecasting strategy employed by each policy determines whether a replacement is profitable (i.e., it leads to a lower usage cost). Note that this approach allows multiple copies of contents at different SCBSs.

\begin{figure*}[t]
\centering
\includegraphics[width=0.8\textwidth]{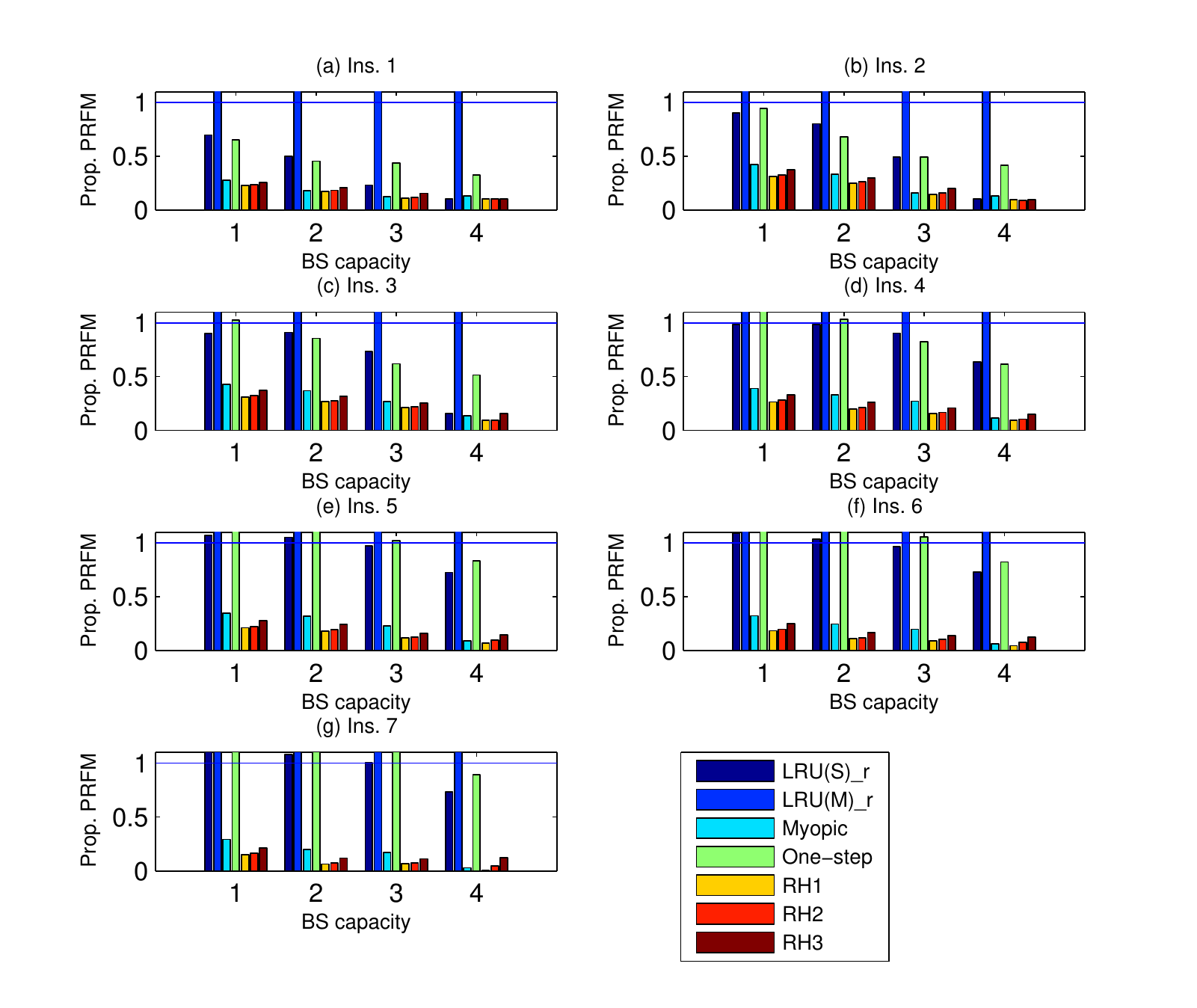}
\caption{Numerical results for online model when replacement heuristic is used on all DP policies.}
\label{fig:online_compare2}
\end{figure*}

The results of this evaluation are summarized in Fig. \ref{fig:online_compare2}. 
By comparing these results with those in Table \ref{tb:resultsOnline}, where Bellman equation is solved exactly, we can see that the performance of the dynamic programming policies changes slightly. The usage cost for $LRU(M)_r$ and $LRU(S)_r$ policies is the same in both comparisons, as the replacement heuristic is only implemented on dynamic programming policies. We can observe that in some cases the replacement heuristic gives even slightly smaller overall cost. 
Further, by comparing the results of the dynamic programming policies with the ones shown in Fig. \ref{fig:results_online_removey}
we note that the replacement heuristic improves the performance in all $*.4$ scenarios significantly, which is attributed to the relaxation of the one-copy per-content condition.  Also, from Fig. \ref{fig:online_compare2} 
we observe that similarly to the previous subsections, the proposed schemes RH1-RH2 achieve the minimal usage cost. More importantly, the performance of the RH policies improve with the size of the studied problem (larger network and larger number of contents). For example, for \textit{Ins.7}, which corresponds to a topology consisting of 15 SCBSs and 1000 contents, the RH1 cost is only 0.04\% higher than the theoretical lower bound, where the latter cannot be achieved in practice since the replacement penalty in this case is zero (assume no repay when replacing a cached content by another). This improved performance on larger networks justifies the reliability of RH policies in practice. 
Also, the exact number of cache updates (replacements) one should allow does not have to be optimized. The number of maximum replacements always can be set to a large number, as in practice a replacement decision is made only when it reduces the immediate cost together with a short prediction in the future that is captured in \ref{eq:rollinghorizon}. Nevertheless, the computation time is largely reduced with the heuristic. For larger instances, e.g., \textit{Ins.7}, for every stage (one-hour interval) the cache update decision can be made in approximately 1 second. 

\section{Discussion and conclusions}
\label{sec:conclusions}

In this paper, we presented a \textit{rolling horizon} cache optimization scheme in a network of SCBSs which collaboratively optimize cache updates.
From the evaluation, we note that even for low per-hour content generation rates, the gains of the proposed scheme exceed 69-99\% the performance of the offline schemes. In order to reduce the complexity of the online scheme, we approximate future content updates/replacements cost by considering only a limited history horizon. To further reduce the time needed to solve the online cache optimization problem we propose two simplifications to the problem: (a) we restrict the number of content replicas in the network and (b) we limit the allowed content replacements. The experimental results show that policies achieving the higher cache hit ratio do not necessarily coincide with the ones with the minimal usage cost. Also, the results make clear the value of considering future information when deciding the cache update policy. When cache capacity is limited, restricting the number of content replicas is shown to be very efficient. Limiting the allowed content updates, in general, leads to easy-to-deploy schemes especially when the \textit{rolling horizon} approximation is used. Nevertheless, the size of the horizon depends on the computational capacity and the inaccuracy of forecasting we can afford. 

\bibliographystyle{IEEEtran}

\begin{thebibliography}{10}
\providecommand{\url}[1]{#1}
\csname url@samestyle\endcsname
\providecommand{\newblock}{\relax}
\providecommand{\bibinfo}[2]{#2}
\providecommand{\BIBentrySTDinterwordspacing}{\spaceskip=0pt\relax}
\providecommand{\BIBentryALTinterwordstretchfactor}{4}
\providecommand{\BIBentryALTinterwordspacing}{\spaceskip=\fontdimen2\font plus
\BIBentryALTinterwordstretchfactor\fontdimen3\font minus
  \fontdimen4\font\relax}
\providecommand{\BIBforeignlanguage}[2]{{%
\expandafter\ifx\csname l@#1\endcsname\relax
\typeout{** WARNING: IEEEtran.bst: No hyphenation pattern has been}%
\typeout{** loaded for the language `#1'. Using the pattern for}%
\typeout{** the default language instead.}%
\else
\language=\csname l@#1\endcsname
\fi
#2}}
\providecommand{\BIBdecl}{\relax}
\BIBdecl

\bibitem{Cisco}
``{Global Mobile Data Traffic Forecast Update, 2014--2019},'' Feb. 2015, {Cisco
  Visual Networking Index. Cisco Inc.}

\bibitem{VakaliIntComp2003}
A.~Vakali and G.~Pallis, ``{Content Delivery Networks: Status and Trends},''
  \emph{{IEEE Internet Computing}}, vol.~7, no.~6, pp. 68--74, Dec. 2003.

\bibitem{Akamai}
``{State of the Internet Report},'' Q2 2014, {[Online]. Available:
  http://www.akamai.com/, Akamai Inc.}

\bibitem{KnapsackProbBook}
S.~Martello and P.~Toth, \emph{{Knapsack Problems}}.\hskip 1em plus 0.5em minus
  0.4em\relax New York: Wiley, 1990.

\bibitem{GolrezaeiCommMag2013}
N.~Golrezaei, A.~F. Molisch, A.~G. Dimakis, and G.~Caire, ``{Femtocaching and
  Device-to-Device Collaboration: A New Architecture for Wireless Video
  Distribution},'' \emph{{IEEE Communications Magazine}}, vol.~51, no.~4, pp.
  142--149, Apr. 2013.

\bibitem{GolrezaeiTITApr2014}
N.~Golrezaei, A.~G. Dimakis, and A.~F. Molisch, ``{Scaling Behavior for
  Device-to-Device Communications With Distributed Caching},'' \emph{IEEE
  Trans. Information Theory}, vol.~60, no.~7, pp. 4286--4298, Jul. 2014.

\bibitem{Shokrollahi06}
A.~Shokrollahi, ``{Raptor codes},'' \emph{IEEE Trans. Information Theory},
  vol.~52, no.~6, pp. {2551--2567}, June 2006.

\bibitem{AhlswedeTIT00}
R.~Ahlswede, N.~Cai, S.-Y.~R. Li, and R.~W. Yeung, ``{Network Information
  Flow},'' \emph{IEEE Trans. Information Theory}, vol.~46, no.~4, pp.
  1204--1216, Jul. 2000.

\bibitem{MaddahAliTIT2014}
M.~A. Maddah-Ali and U.~Niesen, ``{Fundamental Limits of Caching},'' \emph{IEEE
  Trans. Information Theory}, vol.~60, no.~5, pp. 2856--2867, May 2014.

\bibitem{GregoriJSAC2016}
M.~Gregori, J.~G. Vilardebo, J.~Matamoros, and D.~Gunduz, ``{Wireless Content
  Caching for Small Cell and D2D Networks},'' \emph{IEEE Journal on Selected
  Areas in Communications}, vol.~34, no.~5, pp. 1222--1234, May 2016.

\bibitem{PoularakisTCOM2014}
K.~Poularakis, G.~Iosifidis, and L.~Tassiulas, ``{Approximation Algorithms for
  Mobile Data Caching in Small Cell Networks},'' \emph{IEEE Trans.
  Communications}, vol.~62, no.~10, pp. 3665--3677, Oct. 2014.

\bibitem{PoularakisInfocom16}
K.~Poularakis, G.~Iosifidis, A.~Argyriou, I.~Koutsopoulos, and L.~Tassiulas,
  ``{Caching and Operator Cooperation Policies for Layered Video Content
  Delivery},'' in \emph{{Proc. IEEE INFOCOM 2016}}, San Francisco, CA, USA,
  Apr. 2016.

\bibitem{KhreishahJSAC2017}
A.~Khreishah, J.~Chakareski, and A.~Gharabeih, ``{Joint Caching, Routing, and
  Channel Assignment for Collaborative Small-Cell Cellular Networks},''
  \emph{IEEE Journal on Selected Areas in Communications}, vol.~34, no.~8, pp.
  2275--2284, Aug. 2016.

\bibitem{ThomosTMM2015}
N.~Thomos, E.~Kurdoglu, P.~Frossard, and M.~V. der Schaar, ``{Adaptive
  Prioritized Random Linear Coding and Scheduling for Layered Data Delivery
  from Multiple Servers},'' \emph{{IEEE Trans. on Multimedia}}, vol.~17, no.~6,
  pp. 893--906, Jun. 2015.

\bibitem{MaggiCacheComCom2018}
L.~Maggi, L.~Gkatzikis, G.~Paschos, and J.~Leguay, ``{Adapting Caching to
  Audience Retention Rate},'' \emph{Computer Communications}, vol. 116, pp.
  159--171, Jan. 2018.

\bibitem{FamaeyJNCA2013}
J.~Famaey, F.~Iterbeke, T.~Wauters, and F.~D. Turck, ``{Towards a Predictive
  Cache Replacement Strategy for Multimedia Content},'' \emph{{Elsevier Journal
  of Networks and Computer Applications}}, vol.~36, no.~1, pp. 219--227, Jan.
  2013.

\bibitem{LiTMM17}
S.~Li, J.~Xu, M.~van~der Schaar, and W.~Li, ``{Trend-Aware Video Caching
  Through Online Learning},'' \emph{IEEE Trans. Multimedia}, vol.~18, no.~12,
  pp. 2503--2516, Dec. 2016.

\bibitem{ZhangICCW18}
N.~Zhang, K.~Zheng, and M.~Tao, ``{Using Grouped Linear Prediction and
  Accelerated Reinforcement Learning for Online Content Caching},'' in
  \emph{{Proc. of IEEE Int. Conf. on Communications Workshops, ICCWÕ18}},
  Kansas City, MO, USA, May 2018.

\bibitem{YangArxiv18}
P.~Yang, N.~Zhang, S.~Zhang, L.~Yu, J.~Zhang, and X.~Shen, ``{Content
  Popularity Prediction Towards Location-Aware Mobile Edge Caching},''
  \emph{{available at https://arxiv.org/abs/1809.00232}}, Sep. 2018.

\bibitem{AbadArxiv18}
M.~S.~H. Abad, E.~Ozfatura, O.~Ercetin, and D.~Gunduz, ``{Dynamic Content
  Updates in Heterogeneous Wireless Networks},'' \emph{{available at
  https://arxiv.org/abs/1902.09445}}, Feb. 2019.

\bibitem{SadeghiJTSP18}
A.~Chattopadhyay, B.~Blaszczyszyn, and H.~P. Keeler, ``{Optimal and Scalable
  Caching for 5G Using Reinforcement Learning of Space-Time Popularities},''
  \emph{{IEEE Journal on Selected Topics on Signal Processing}}, vol.~12,
  no.~1, pp. 180--190, Feb. 2018.

\bibitem{NegliaToN18}
G.~Neglia, D.~Carra, and P.~Michiardi, ``{Cache Policies for Linear Utility
  Maximization},'' \emph{{IEEE/ACM Trans. on Networking}}, vol.~26, no.~1, pp.
  302--313, Feb. 2018.

\bibitem{BharathTCOM18}
B.~N. Bharath, K.~G. Nagananda, D.~Gunduz, and H.~V. Poor, ``{Caching With
  Time-Varying Popularity Profiles: A Learning-Theoretic Perspective},''
  \emph{{IEEE Trans. on Communications}}, vol.~66, no.~8, pp. 3837--3847, Sep.
  2018.

\bibitem{ChattopadhyayTWC18}
A.~Chattopadhyay, B.~Blaszczyszyn, and H.~P. Keeler, ``{Gibbsian On-Line
  Distributed Content Caching Strategy for Cellular Networks},'' \emph{{IEEE
  Trans. on Wireless Communications}}, vol.~17, no.~2, pp. 969--981, Feb. 2018.

\bibitem{GharaibehInfocom16}
A.~Gharaibeh, A.~Khreishah, and I.~Khalil, ``{An O(1)-Competitive Online
  Caching Algorithm for Content Centric Networking},'' in \emph{{Proc. IEEE
  INFOCOM}}, San Francisco, CA, USA, Apr. 2016.

\bibitem{MullerTWC2017}
S.~Muller, O.~Atan, M.~van~der Schaar, and A.~Klein, ``{Context-Aware Proactive
  Content Caching With Service Differentiation in Wireless Networks},''
  \emph{{Trans. on Wireless Communications}}, vol.~16, no.~2, pp. 1024--1036,
  Feb. 2017.

\bibitem{SaltarinNetworking2018}
J.~Saltarin, T.~Braun, E.~Bourtsoulatze, and N.~Thomos, ``{PopNetCod: A
  Popularity-based Caching Policy for Network Coding enabled Named Data
  Networking},'' in \emph{{Proc. of IFIP Networking Conference,
  NetworkingÕ18}}, Zurich, Switzerland, May 2018.

\bibitem{GharaibehTPDS2016}
A.~Gharaibeh, A.~Khreishah, I.~Khalil, and J.~Wu, ``{Distributed Online
  En-route Caching},'' \emph{{IEEE Trans. on Parallel and Distributed
  Systems}}, vol.~27, no.~12, pp. 3455--3468, Dec. 2016.

\bibitem{KhreishahInfocomWK2015}
A.~Khreishah and J.~Chakareski, ``{Collaborative Caching for
  Multicell-Coordinated Systems},'' in \emph{{Proc. IEEE INFOCOM Workshop
  Communication and Networking Techniques for Contemporary Video}}, Hong Kong,
  Apr. 2015.

\bibitem{GharaibehTMC2016}
A.~Gharaibeh, A.~Khreishah, B.~Ji, and M.~Ayyash, ``{A Provably Efficient
  Online Collaborative Caching Algorithm for Multicell-Coordinated Systems},''
  \emph{{IEEE Trans. on Mobile Computing}}, vol.~15, no.~8, pp. 1863--1876,
  Aug. 2016.

\end{thebibliography}

\end{document}